\documentstyle[aps,prd,epsfig,preprint]{revtex}
\def\fun#1#2{\lower3.6pt\vbox{\baselineskip0pt\lineskip.9pt
\ialign{$\mathsurround=0pt#1\hfill##\hfil$\crcr#2\crcr\sim\crcr}}}
\begin{document}
\vspace{0.5in}
\title{\vskip-2.5truecm{\hfill \baselineskip 14pt 
{\hfill {{\small        \hfill UT-STPD-4/98 }}}\\
{{\small        \hfill BA-98-11}}\vskip .1truecm} 
\vspace{1.0cm}
\vskip 0.1truecm {\bf R-Symmetry in the Minimal Supersymmetric 
Standard Model and Beyond with Several Consequences }}
\vspace{1cm}
\author{{G. Lazarides}$^{(1)}$\thanks{lazaride@eng.auth.gr} 
{and Q. Shafi}$^{(2)}$\thanks{shafi@bartol.udel.edu}} 
\vspace{1.0cm}
\address{$^{(1)}${\it Physics Division, School of Technology, 
Aristotle University of Thessaloniki,\\ Thessaloniki GR 540 06, 
Greece.}}
\address{$^{(2)}${\it  Bartol Research Institute, University of 
Delaware,\\
Newark, DE 19716, USA}}
\maketitle

\vspace{2cm}

\begin{abstract}
\baselineskip 12pt

\par
The supersymmetric sector of minimal supersymmetric standard model 
(MSSM) possesses a $U(1)$ R-symmetry which contains $Z_2$ matter 
parity. Non-zero neutrino masses, consistent with a `redefined' 
R-symmetry, are possible through the see-saw mechanism and/or a pair 
of superheavy (mass $M$) $SU(2)_L$ triplets with vev $\sim M^2_W/M$. 
If this R-symmetry is respected by the higher order terms, then baryon 
number conservation follows as an immediate consequence. In the presence 
of right handed neutrinos, the observed baryon asymmetry of the universe 
arises via leptogenesis. An interplay of R- and Peccei-Quinn symmetry 
simultaneously resolves the strong CP and $\mu$ problems. 

\end{abstract}

\thispagestyle{empty}
\newpage
\pagestyle{plain}
\setcounter{page}{1}
\def\beq{\begin{equation}}
\def\eeq{\end{equation}}
\def\beqa{\begin{eqnarray}}
\def\eeqa{\end{eqnarray}}
\def\tr{{\rm tr}}
\def\x{{\bf x}}
\def\p{{\bf p}}
\def\k{{\bf k}}
\def\z{{\bf z}}
\baselineskip 20pt

\par
Although quite compelling, the minimal supersymmetric standard model 
(MSSM) fails to address a number of important challenges. For 
instance, to explain the apparent stability of the proton, it must be 
assumed that the dimensionless coefficients accompanying dimension five 
operators are of order $10^{-8}$ or less. The strong CP and $\mu$ 
problems loom large in the background, and the observed baryon asymmetry, 
it appears, cannot be explained within the MSSM framework. Last, but by 
no means least, there is increasing evidence for non-zero neutrino masses 
from a variety of experiments.

\par
In a recent paper\cite{dls}, we offered one approach for resolving many
of the above problems. It relied on extending the gauge symmetry to
$SU(2)_L \times SU(2)_R \times U(1)_{B-L}$ , with a global $U(1)$ 
R-symmetry playing an essential role. The magnitude of the supersymmetric 
$\mu$ term of MSSM was directly related to the gravitino mass $m_{3/2}
~(\sim 1~{\rm{TeV}})$ which, in turn, arises from the hidden sector 
{\it a la} supergravity. The left-right symmetry ensures the presence of 
right handed neutrino superfields and consequently non-zero neutrino 
masses, while the R-symmetry implies an accidental global $U(1)_B$ 
symmetry which explains why the proton is so stable. Note that the 
R-symmetry is spontaneously and perhaps even explicitly broken by the 
hidden sector. The soft (quadratic and trilinear) supersymmetry breaking 
terms in the visible sector are expected to explicitly break the 
R-symmetry.

\par
In this paper, we wish to provide a resolution of the problems listed  
above without departing  from the $SU(3)_c \times SU(2)_L 
\times U(1)_Y$ framework of MSSM. We observe that the MSSM 
superpotential $W$ possesses a global $U(1)$ R-symmetry\cite{hr} in 
which $Z_2$ matter parity is embedded. We show how neutrino masses can 
be incorporated while preserving a (redefined) R-symmetry. When extended 
to higher orders, this symmetry ensures the appearance of global $U(1)_B$, 
thereby guaranteeing proton stability. In the case where right handed 
neutrinos are included, the observed baryon asymmetry of the universe can 
arise, as we will see, via leptogenesis. The approach followed here also 
provides the framework for an elegant resolution of the strong CP and 
$\mu$ problems of MSSM, with the R-symmetry once again playing an 
essential role. 

\par
The MSSM superpotential $W$ contains the following renormalizable terms 
(we will not distinguish between the generations in this paper):
\begin{equation}
H^{(1)}QU^c,~H^{(2)}QD^{c},~H^{(2)}LE^{c},~H^{(1)} H^{(2)}~.
\label{eq:superpot}
\end{equation}
Here $H^{(1)},~H^{(2)}$ are the two higgs superfields, $Q$ denotes the 
$SU(2)_L$ doublet quark superfields, $U^c$ and $D^c$ are the $SU(2)_L$ 
singlet quark superfields, while $L~(E^c)$ stands for the $SU(2)_L$ 
doublet (singlet) lepton superfields. A $Z_2$ matter parity 
($Z_2^{mp}$) under which only the `matter' superfields change sign 
ensures the absence of terms such as $QD^{c}L$ and $U^cD^cD^c$ which 
lead to rapid proton decay.

\par
The superpotential in Eq.(\ref{eq:superpot}) possesses three global 
symmetries, namely, $U(1)_B,~U(1)_L$ and $U(1)_R$. (Except for 
sphaleron effects in baryogenesis, we will ignore the `tiny' 
non-perturbative violation of $B$ and $L$ by the $SU(2)_L$ instantons.) 
The `global' charges of the various superfields are as follows:
\begin{eqnarray*}
B:~H^{(1)}(0),~H^{(2)}(0),~Q(1/3),~U^{c}(-1/3),~D^{c}(-1/3),
~L(0),~E^{c}(0)~;
\end{eqnarray*}
\begin{eqnarray*}
L:~H^{(1)}(0),~H^{(2)}(0),~Q(0),~U^c(0),~D^c(0),
~L(1),~E^c(-1)~;
\end{eqnarray*}
\begin{equation}
R:~H^{(1)}(1),~H^{(2)}(1),~Q(1/2),~U^c(1/2),~D^c(1/2),
~L(1/2),~E^c(1/2)~.
\label{eq:sym}
\end{equation}
We have normalized the $R$ charges such that $W$ carries two units.

\par
The introduction of the right handed neutrino superfields, $\nu^c$,
gives rise, consistent with $Z_2^{mp}$, to two additional 
renormalizable superpotential couplings
\begin{equation}
H^{(1)}L\nu^c ~,~M^{R} \nu^c \nu^c~,
\label{eq:nent}
\end{equation}
where $M^{R}$ is the Majorana mass matrix of the superheavy right 
handed neutrinos. The first term in Eq.(\ref{eq:nent}) fixes the 
quantum numbers of the right handed neutrinos, namely, $B(\nu^c)=0$, 
$L(\nu^c)=-1$, $R(\nu^c)= 1/2$. The second term violates both 
$U(1)_L$ and $U(1)_R$, but the combination
\begin{equation}
R^\prime = R - \frac {1}{2}~L
\label{eq:rsym}
\end{equation}
is now the new R-symmetry of the superpotential. In addition, 
the $Z^{lp}_2$ (lepton parity) subgroup of $U(1)_L$, under which 
only the lepton superfields $L,~E^c$ change sign, remains unbroken. 
Consequently, the global symmetries of the renormalizable superpotential 
containing all the couplings in Eqs.(\ref{eq:superpot}) and 
(\ref{eq:nent}) are $U(1)_{R^{\prime}}$ , $U(1)_B$ and 
$Z^{lp}_2$. With the couplings in Eq.(\ref{eq:nent}), the observed 
neutrinos acquire masses via the well-known see-saw mechanism\cite{grs}.

\par
It is interesting to note that both $U(1)_B$ and lepton parity are 
automatically implied by $U(1)_{R^{\prime}}$. Moreover, this remains 
true even if non-renormalizable terms are included in the superpotential. 
Indeed, by extending the $U(1)_{R^\prime}$ symmetry to higher order 
terms, we will first show that $U(1)_B$ follows as a consequence. To see 
this, note that $U(1)_{R^{\prime}}$ contains $Z_2^{bp}$ (baryon 
parity) under which only the color triplet, antitriplet $(3,~\bar{3})$ 
superfields change sign. This means that superpotential couplings 
containing, in addition to color singlet and $(3 \cdot \bar{3})^m$ 
(m $\geq$ 0) factors, the $U(1)_B$ violating combinations 
$(3 \cdot 3 \cdot 3)^n$ or $(\bar{3} \cdot \bar{3} \cdot 
\bar{3})^n$ with $n$ = odd $\geq 1$ are not allowed. Similarly, 
analogous couplings but with $n$ = even $\geq 2$ are also not allowed 
since their $R^\prime$ charge exceeds two units and cannot be 
compensated. In particular, the troublesome dimension five operators 
$QQQL$ and $U^cU^cD^cE^c$ are eliminated.

\par
One can next show that $U(1)_{R^{\prime}}$ implies $Z^{lp}_2$
(lepton parity) to all orders. Because of $U(1)_B$, the quark
superfields must appear in `blocks' $QU^c(1)$ and $QD^c(1)$, where 
the parenthesis indicates the $R^\prime$ charge. The other 
non-leptonic `blocks' are $H^{(1)}(1)$ and $H^{(2)}(1)$. The leptonic 
superfields are $L(0),~E^c(1),~\nu^c(1)$. To violate lepton parity, 
we need an odd number of lepton superfields. Therefore, we should consider:
i) odd number of $L$ 's together, by $U(1)_{R^\prime}$ symmetry, with
two non-leptonic blocks belonging to the four types described above;
ii) even number of $L$ 's and a single $E^c$ or $\nu^c$, together with
one non-leptonic block;
iii) odd number of $L$ 's with two out of the $E^c$ 's and $\nu^c$ 's. 
In all three cases, one ends up with an odd number of $SU(2)_L$ doublets 
which is not gauge invariant. 

\par
In summary, both lepton parity and $U(1)_B$ 
are present in the scheme to all orders as mere 
consequences of the $U(1)_{R^{\prime}}$ symmetry and remain exact, 
although $U(1)_{R^{\prime}}$ is explicitly broken to its maximal 
non-R-subgroup $Z_4^{\prime}$ (which includes $Z_2^{bp}$) by the 
supersymmetry breaking terms in the visible sector. 

\par
We now present an alternative scheme for introducing non-zero neutrino 
masses in MSSM . This scheme is, actually, quite familiar\cite{w} from 
Grand Unified Theories (GUTs), and was recently considered within the 
non-supersymmetric standard model framework in Ref.\cite{ms}. Introduce, 
in MSSM, an $SU(2)_L$ triplet pair $T,~\bar{T}$, with hypercharges +1, 
-1 respectively. Consider the renormalizable superpotential couplings
\begin{equation}
TLL,~\bar{T} H^{(1)} H^{(1)},~MT\bar{T}~,
\label{eq:triplet}
\end{equation}
such that $B(T)=B(\bar{T})=0$, $L(T)=-2$, $L(\bar{T})=0$, $R(T)=1$, 
$R(\bar{T})=0$, from the first two couplings, and $M$ is some 
superheavy scale (taken real and positive by suitable phase redefinitions
of the superfields $T$, $\bar{T}$). The supersymmetric mass term in 
Eq.(\ref{eq:triplet}) breaks $U(1)_R$ and $U(1)_L$ but, in analogy with 
the previous discussion involving the $\nu^c$ superfields, the 
superpotential defined by the terms in Eqs.(\ref{eq:superpot}) and 
(\ref{eq:triplet}) possesses a redefined R-symmetry generated by
\begin{equation}
R^{\prime \prime} = R - \frac {1}{2}~L~.
\label{eq:rrsym}
\end{equation} 
The $R^{\prime \prime}$ charges of the various superfields are:
\begin{equation}
H^{(1)}(1),~H^{(2)}(1),~Q(1/2),~U^c(1/2),~D^c(1/2),
~L(0),~E^{c}(1),~T(2),~\bar{T}(0)~.
\label{eq:charge}
\end{equation}
Both $U(1)_B$ and lepton parity remain unbroken in this case too. 
Finally, as with the $U(1)_{R^\prime}$ symmetry, one can readily 
show that $U(1)_{R^{\prime \prime}}$ implies conservation of $B$ 
and $Z^{lp}_{2}$ to all orders despite its explicit breaking to its 
maximal non-R-subgroup $Z_4^{\prime\prime}$ (including $Z_2^{bp}$) 
by the supersymmetry breaking terms in the visible sector.

\par
It is readily checked that the scalar component of $T$ 
acquires a non-zero vacuum expectation value (vev) 
$\sim M^2_W/M~(\ll M_W)$, with the electroweak breaking 
playing an essential role in the generation of this vev. This is due to 
the fact that the two last terms in Eq.(\ref{eq:triplet}), after 
electroweak breaking, give rise to a term linear with respect to $T$ in 
the scalar potential of the theory. The vev of $T$ leaves $U(1)_B$ , 
$Z^{lp}_{2}$ and the $Z_4^{\prime\prime}$ subgroup of 
$U(1)_{R^{\prime\prime}}$ unbroken and generates non-zero neutrino 
masses. Note that $T$, $\bar{T}$ are superheavy fields, so that the 
low energy spectrum is given by the MSSM.

\par
It should be clear that the coexistence of all the superpotential couplings 
in Eqs.(\ref{eq:superpot}), (\ref{eq:nent}) and (\ref{eq:triplet})
provides us with a scheme where the light neutrino masses acquire 
contributions from the see-saw mechanism as well as the triplet vev. 
It is important to note that, in this `combined' case, 
$W$ possesses a $U(1)$ R-symmetry, $U(1)_{\hat{R}}$ , 
which coincides with $U(1)_{R^{\prime}}$ or $U(1)_{R^{\prime\prime}}$ 
when restricted to the superfields where these symmetries are defined.
(Note that $U(1)_{R^{\prime}}$ and $U(1)_{R^{\prime\prime}}$ 
become identical when restricted to the MSSM superfields.) 
This R-symmetry, just as in the previous cases, implies $U(1)_B$ and 
$Z^{lp}_{2}$ to all orders. It is, finally, interesting to notice that 
baryon number and lepton parity conservation is a consequence of 
the `redefined' R-symmetries in Eqs.(\ref{eq:rsym}), (\ref{eq:rrsym}) 
or $U(1)_{\hat{R}}$ , in the `combined' case, and not of the original 
$U(1)_R$ which allows couplings like $QQQL$ and $U^cU^cD^cE^c$.

\par
The two mechanisms considered above for generating masses for the 
neutrinos have an additional far  reaching consequence. This has to do 
with the generation of the observed baryon asymmetry in the universe.
The basic idea is to generate an initial lepton asymmetry which is 
partially transformed through the non-perturbative electroweak sphaleron 
effects, that `actively' violate $B+L$ at energies above $M_W$, to the 
observed baryon asymmetry. Actually, this is the only way to generate 
baryons in the present scheme, since baryon number is otherwise 
exactly conserved. This mechanism has been well documented\cite{lepto} 
when the lepton asymmetry is created by a decaying massive Majorana 
neutrino (say from the $\nu^c$ superfields) and exploits the couplings 
given in Eq.(\ref{eq:nent}). If the $T,~\bar {T}$ superfields with 
the couplings given in Eq.(\ref{eq:triplet}) are also present, then 
additional diagrams must be considered. 

\par
\begin{figure}
\epsfig{figure=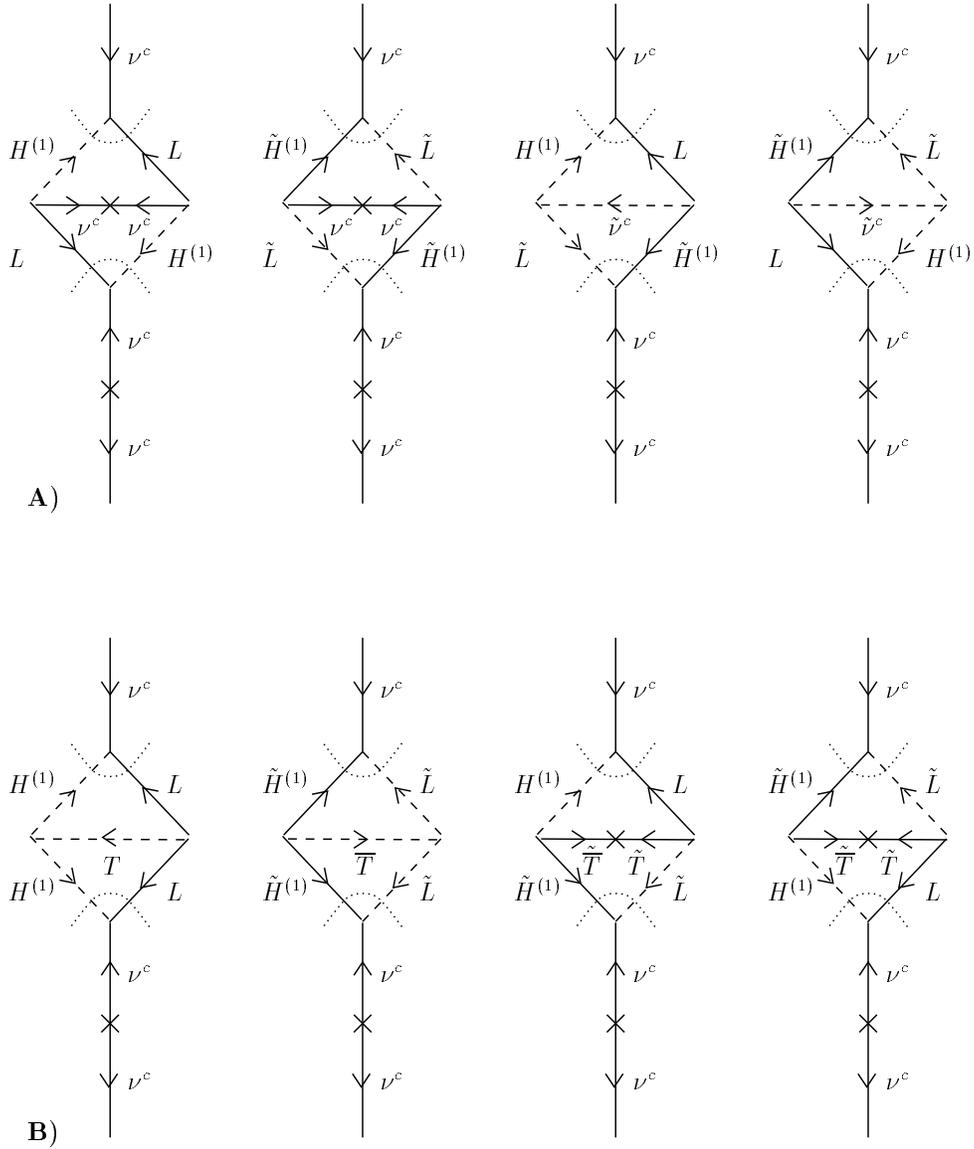,height=7in}
\medskip
\caption{The double-cut diagrams for leptogenesis via the decay of 
fermionic $\nu^c$ 's. The diagrams A (B) correspond to the exchange 
of $\nu^c$ ($T,~\bar{T}$) superfields. Continuous (dashed) lines 
represent fermions (bosons) while tildes denote the supersymmetric 
partners.
\label{fig:graph}}
\end{figure}

The complete set of double-cut diagrams for leptogenesis from a 
decaying fermionic $\nu^c$, which is the relevant case for 
inflationary models where the inflaton predominantly decays to a 
fermionic right handed neutrino, is displayed in Fig.\ref{fig:graph}. 
The resulting lepton asymmetry, in this case, can be estimated\cite{lss} 
to be
\begin{equation}
\frac {n_{L}}{s} \approx  \frac{3}{16 \pi}~\frac {T_{r}}
{m_{infl}}~M^{R}_{i}~\frac{{\rm{Im}}(M^{D}~m^{\dagger}
~\tilde{M}^{D})_{ii}}{|\langle H^{(1)} \rangle |^{2}
(M^{D}~M^{D}\,^{\dagger})_{ii}}~\cdot
\label{eq:genlept}
\end{equation}
Here $T_{r}$ is the `reheat' temperature, $m_{infl}$ the inflaton 
mass, $M^{D}$ the neutrino `Dirac' mass matrix in the basis where 
the Majorana mass matrix of right handed neutrinos, $M^R$, is diagonal 
with positive entries, and $M^{R}_{i}$ is the mass of the decaying 
$\nu^{c}_{i}$. Also
\begin{equation}
m \approx -\alpha~ t~ \frac{\langle H^{(1)}\rangle^{2}}{M}
-\tilde{M}^{D}\ \frac{1}{M^{R}}\ M^{D}~ 
\label{eq:mass}
\end{equation}
is the light neutrino mass matrix in the same basis, with $t$ (a complex 
symmetric matrix) and $\alpha$ being the coefficients of the first and 
second terms in Eq.(\ref{eq:triplet}) respectively. It should be noted 
that this estimate holds provided that $M^{R}_i$ is much smaller than 
the mass of the other $\nu^c$ 's and the mass $M$ of the triplets . 
Eq.(\ref{eq:genlept}) gives\cite{lss} the bound
\begin{equation}
\left |{n_L\over s}\right| \stackrel{_<}{_\sim}{3\over 16 \pi} 
~{T_r\over m_{infl}}~{M^{R}_{i}~m_{\nu_{\tau}}\over |\langle 
H^{(1)}\rangle|^2}~, 
\label{eq:bound}
\end{equation}
which, for $T_{r}\approx 10^{9}~{\rm{GeV}}$ (consistent with 
the gravitino constraint), $m_{infl}\approx 3\times 10^{13}
~{\rm{GeV}}$, $M^{R}_{i}\approx 10^{10}~{\rm{GeV}}$ 
(see Ref.\cite{lss}), $|\langle H^{(1)}\rangle|\approx 
174~{\rm{GeV}}$, and  $m_{\nu_{\tau}}\approx 5~{\rm{eV}}$ 
(providing the hot dark matter of the universe), gives 
$|n_L/s|\stackrel{_<}{_\sim} 3 \times 10^{-9}$. This is large 
enough to account for the observed baryon asymmetry of the universe.
It is important though to ensure that the lepton asymmetry is not 
erased by lepton number violating $2\rightarrow 2$ scatterings at 
all temperatures between $T_r$ and 100 GeV. This requirement
gives\cite{ibanez} $m_{\nu_{\tau}}\stackrel{_<}{_\sim} 
10~{\rm{eV}}$.

\par
We pointed out that, in non-inflationary (and perhaps some 
inflationary) models, leptogenesis from the decay of bosonic $\nu^c$ 's 
as well as  bosonic and fermionic $T$, $\bar{T}$ 's may be present too. 
Most of the relevant double-cut diagrams can be obtained from the ones 
in Fig.\ref{fig:graph} by breaking up the $\nu^c$, $T$, $\bar{T}$ 
internal lines and joining the external $\nu^c$ lines.
The only extra diagram (not obtainable this way) is a diagram of the same 
type with bosonic $\nu^c$ external lines and a fermionic $T$, $\bar{T}$ 
internal line. The important observation is that diagrams of the type in
Fig.\ref{fig:graph} with no $\nu^c$ internal or external lines cannot be 
constructed. Thus, efficient leptogenesis can take place only in the 
presence of $\nu^c$ 's.

\par
We have seen how $U(1)_B$ arises as a consequence of requiring the 
superpotential $W$ (including higher order terms) to respect a 
$U(1)$ R-symmetry. Among other things, this explains why the proton 
is so stable. However, the learned reader may be concerned that 
requiring the non-renormalizable terms in the superpotential to respect 
a continuous R-symmetry may not be a reasonable thing to do. Indeed, one 
may wonder if continuous global symmetries such as $U(1)_{\hat{R}}$ 
or the Peccei-Quinn\cite{pq} ($U(1)_{PQ}$) symmetry, rather than being 
imposed, can arise in some more `natural' manner. One way how this may 
occur was pointed out in Ref.\cite{lps}. Here, discrete (including R-) 
symmetries that typically arise after compactification could effectively 
behave as if they are continuous. Furthermore, such `continuous' symmetries 
can be very useful in resolving problems other than the one of 
proton stability. To see this, let us now address the strong CP and 
$\mu$ problems of MSSM. It has been noted by earlier authors\cite{np}
that a continuous $U(1)$ R-symmetry can be relevant for the solution 
of the $\mu$ problem. By invoking  $U(1)_{PQ}$ and combining it with 
the $U(1)$ R-symmetry above, we will provide a resolution of both the 
strong CP and $\mu$ problems, with the $U(1)$ R-symmetry playing an 
essential role in controlling the structure of the terms that are 
permitted at the non-renormalizable level.

\par
It has been recognized for some time that, within the supergravity 
extension of MSSM, the existence of D- and F-flat directions in 
field space can generate an intermediate scale $M_I$ which, 
in the simplest case, is given by
\begin{equation}
M_I \sim \sqrt{m_{3/2}M_P} \sim 10^{11}-10^{12}~{\rm{GeV}}~, 
\label{eq:inter}
\end{equation}
where $m_{3/2} \sim 1~{\rm{TeV}}$ is the supersymmetry breaking 
scale and $M_P = 1.22 \times 10^{19}$ GeV is the Planck mass. 
It seems `natural' to try and identify $M_I$ with the symmetry 
breaking scale $f_{a}$ of $U(1)_{PQ}$, such that 
$\mu \sim m_{3/2} \sim f^2_{a}/M_P$\cite{kn}. We will 
now see how this idea, which simultaneously resolves the strong CP 
and $\mu$ problems, can be elegantly realized in the presence of the
$U(1)$ R-symmetry. Note that the resolution of the $\mu$ problem 
forces us to consider non-renormalizable terms.

\par
We supplement the MSSM spectrum with a pair of superfields 
$N,~\bar{N}$ whose vevs will break $U(1)_{PQ}$ at an 
intermediate scale. $W$ contains\cite{dvali} the following terms:
\begin{equation}
H^{(1)}QU^c,~H^{(2)}QD^c,~H^{(2)}LE^c,~N^{2}H^{(1)}H^{(2)},
~N^{2} \bar{N}^2~.
\label{eq:pquinn}
\end{equation}
The global symmetries of this superpotential are $U(1)_B$, $U(1)_L$ 
(with the new superfields $N$, $\bar{N}$ being neutral under both), 
an anomalous Peccei-Quinn symmetry $U(1)_{PQ}$, and a non-anomalous 
R-symmetry $U(1)_{\tilde{R}}$. The $PQ$ and $\tilde{R}$ charges are 
as follows:
\begin{equation}
\begin{array}{rcl}
PQ:~H^{(1)}(1),~H^{(2)}(1),~Q(-1/2),~U^c(-1/2),~D^c(-1/2),
~~~~~~~~~~\\
~L(-1/2),~E^c(-1/2),~N(-1),~\bar{N}(1)~,
~~~~~~~~~~~~~~~~~~~~\\
& & \\
\tilde{R}:~H^{(1)}(0),~H^{(2)}(0),~Q(1),~U^c(1),~D^c(1),
~L(1),~E^c(1),~N(1),~\bar{N}(0)~.
\end{array}
\label{eq:pqr}
\end{equation}
Note that the quartic terms in Eq.(\ref{eq:pquinn}) carry a 
coefficient proportional to $M^{-1}_P$ which has been left out. 
The R-symmetry ensures that undesirable terms such as $N \bar{N}$, 
which otherwise spoil the flat direction in the supersymmetric limit, 
are absent from Eq.(\ref{eq:pquinn}).

\par
After taking the supersymmetry breaking terms into account, one 
finds\cite{dvali} that, for suitable choice of parameters, a 
solution with
\begin{equation}
|\langle N \rangle|~=~|\langle\bar{N}\rangle|
~\sim \sqrt{m_{3/2}M_{P}}
\label{eq:vev}
\end{equation}
is preferred over the one with 
$\langle N\rangle=\langle\bar{N}\rangle=0$.
To see this, let us consider the relevant part of the scalar potential: 
\begin{equation}
V=\left(m_{3/2}^2+\lambda^2\left|\frac{N\bar{N}}{M_P}\right|^2
\right)(|N|^2+|\bar{N}|^2)+\left(Am_{3/2}\lambda
\frac{N^2\bar{N}^2}{2M_P}+h.c \right)~,
\label{eq:potential}
\end{equation} 
where $\lambda/(2M_P)$ is the coefficient of the last superpotential 
term in Eq.(\ref{eq:pquinn}) and $A$ the dimensionless coefficient of
the corresponding soft supersymmetry breaking term ($\lambda$ is 
taken real and positive by appropriately redefining the phases of 
$N,~\bar{N}$). This potential can be rewritten as
\begin{equation}
V=\left(m_{3/2}^2+\lambda^2\left|\frac{N\bar{N}}{M_P}\right|^2
\right)\left[(|N|-|\bar{N}|)^2+2|N||\bar{N}|\right]+|A|m_{3/2}
\lambda\frac{|N\bar{N}|^2}{M_P}{\rm{cos}}(\epsilon+2\theta
+2\bar{\theta})~,
\label{eq:npotential}
\end{equation}
where $\epsilon,~\theta,~\bar{\theta}$ are the phases of 
$A,~N,~\bar{N}$ respectively. Minimization of $V$ then requires 
$|N|=|\bar{N}|$, $\epsilon+2\theta+2\bar{\theta}=\pi$ and 
$V$ takes the form
\begin{equation}
V=2|N|^2m_{3/2}^2\left(\lambda^2\frac{|N|^4}{m_{3/2}^2M_P^2}-
|A|\lambda\frac{|N|^2}{2m_{3/2}M_P}+1\right)~.
\label{eq:nnpotential}
\end{equation}
It is now obvious that, for $|A|>4$, the absolute minimum of the 
potential is at
\begin{equation}
|\langle N\rangle|=|\langle\bar{N}\rangle|=
(m_{3/2}M_P)^{\frac{1}{2}}
\left(\frac{|A|+(|A|^2-12)^{\frac{1}{2}}}
{6\lambda}\right)^{\frac{1}{2}}\sim (m_{3/2}M_{P})
^{\frac{1}{2}}~.
\label{eq:solution}
\end{equation}
Note that the $\langle N \rangle,~\langle\bar{N}\rangle$ vevs 
together break $U(1)_{\tilde{R}} \times U(1)_{PQ}$ down to $Z_2^{mp}$. 
Substitution of these vevs in Eq.(\ref{eq:pquinn}) shows that the $\mu$ 
parameter of MSSM is of order $m_{3/2}$ as desired.

\par
This discussion is readily extended to include either massive right
handed neutrino superfields $\nu^c$ or the $SU(2)_L$ triplet higgs 
superfields $T,~\bar{T}$. In the $\nu^c$ case, the new terms in the 
superpotential $W$ are $H^{(1)}L\nu^c$ and $N\nu^c\nu^c$. The 
first term yields $B(\nu^c)=0$, $L(\nu^c)=-1$, $PQ(\nu^c)=-1/2$, 
$\tilde{R}(\nu^c)=1$. The second term leaves $U(1)_{B}$ unbroken 
but breaks $U(1)_{L}$, $U(1)_{PQ}$ and $U(1)_{\tilde{R}}$ to a 
`redefined' Peccei-Quinn symmetry $U(1)_{PQ^{\prime}}$ and a `redefined' 
R-symmetry $U(1)_{\tilde{R}^{\prime}}$ with $PQ^{\prime}=PQ-L$ and 
$\tilde{R}^{\prime}=\tilde{R}+PQ-(1/2)L$. Thus, the strong CP problem 
can be resolved. It should be noted that $U(1)_{\tilde{R}^{\prime}}$ 
coincides with $U(1)_{R^\prime}$ in Eq.(\ref{eq:rsym}) when restricted 
to the superfields where $U(1)_{R^\prime}$ is defined. Moreover, just 
as $U(1)_{R^\prime}$, the R-symmetry $U(1)_{\tilde{R}^{\prime}}$ 
contains $Z_2$ baryon parity as a subgroup and implies $U(1)_{B}$ to 
all orders. $Z_2^{mp}$ is contained in $U(1)_{PQ^{\prime}}$ and, 
thus, $Z_2^{lp}$ is also present but not as an automatic consequence 
of $U(1)_{\tilde{R}^{\prime}}$ in this case. 

\par
A similar discussion applies if the triplets $T,~\bar{T}$ are 
introduced in the scheme of Eq.(\ref{eq:pquinn}). The new 
superpotential terms, in this case, are $TLL$, $\bar{T}H^{(1)}H^{(1)}$ 
and $NT\bar{T}$. The first two couplings give $B(T)=B(\bar{T})=0$, 
$L(T)=-2$, $L(\bar{T})=0$, $PQ(T)=1$, $PQ(\bar{T})=-2$, 
$\tilde{R}(T)=0$ and $\tilde{R}(\bar{T})=2$. The last coupling leaves 
unbroken the symmetries $U(1)_{B}$, $U(1)_{PQ^{\prime\prime}}$ and
$U(1)_{\tilde{R}^{\prime\prime}}$ with $PQ^{\prime\prime}=PQ-L$ 
and $\tilde{R}^{\prime\prime}=\tilde{R}+PQ-(1/2)L$. The symmetry 
$U(1)_{\tilde{R}^{\prime\prime}}$ is an extension of 
$U(1)_{R^{\prime\prime}}$ in Eq.(\ref{eq:rrsym}), contains 
$Z_2$ baryon parity and implies $U(1)_{B}$ to all orders. 

\par
Finally, it should be pointed out that $\nu^c$ and $T$, $\bar{T}$ 
can coexist with all the couplings mentioned being present. In this 
`combined' case, $W$ possesses a $U(1)$ Peccei-Quinn (R-) symmetry, 
$U(1)_{\hat{PQ}}$ ($U(1)_{\hat{\tilde{R}}}$), which coincides 
with $U(1)_{PQ^{\prime}}$ ($U(1)_{\tilde{R}^{\prime}})$ or 
$U(1)_{PQ^{\prime\prime}}$ ($U(1)_{\tilde{R}^{\prime\prime}}$) 
when restricted to the superfields where these symmetries are defined. 
The R-symmetry $U(1)_{\hat{\tilde{R}}}$ implies $U(1)_B$ to all orders.
  
\par
We have focused in this paper on MSSM and its extensions, with $Z_2$
matter parity embedded, to begin with, in a $U(1)$ R-symmetry. Non-zero 
neutrino masses, consistent with a redefined $U(1)$ R-symmetry, can be 
introduced in at least two ways. By requiring the higher order 
superpotential couplings to respect this redefined R-symmetry, one can 
i) explain proton stability to be a consequence of an automatic $U(1)_B$ , 
and ii) show that the observed baryon asymmetry can arise via a primordial 
lepton asymmetry provided right handed neutrinos are present. Finally, 
simultaneous resolutions of the strong CP and $\mu$ problems, with 
$\mu \sim f_{a}^2/M_{P}$, can be elegantly accommodated in this scheme. 

\vspace{1cm}

One of us (G. L.) would like to thank G. Dvali for drawing his 
attention to the resolution of the $\mu$ problem via a PQ-symmetry 
and for important discussions on this point. We acknowledge the NATO 
support, contract number NATO CRG-970149. One of us (Q.S.) would also 
like to acknowledge the DOE support under grant number DE-FG02-91ER40626.

\def\ijmp#1#2#3{{ Int. Jour. Mod. Phys. }{\bf #1~}(19#2)~#3}
\def\pl#1#2#3{{ Phys. Lett. }{\bf B#1~}(19#2)~#3}
\def\zp#1#2#3{{ Z. Phys. }{\bf C#1~}(19#2)~#3}
\def\prl#1#2#3{{ Phys. Rev. Lett. }{\bf #1~}(19#2)~#3}
\def\rmp#1#2#3{{ Rev. Mod. Phys. }{\bf #1~}(19#2)~#3}
\def\prep#1#2#3{{ Phys. Rep. }{\bf #1~}(19#2)~#3}
\def\pr#1#2#3{{ Phys. Rev. }{\bf D#1~}(19#2)~#3}
\def\np#1#2#3{{ Nucl. Phys. }{\bf B#1~}(19#2)~#3}
\def\mpl#1#2#3{{ Mod. Phys. Lett. }{\bf #1~}(19#2)~#3}
\def\arnps#1#2#3{{ Annu. Rev. Nucl. Part. Sci. }{\bf
#1~}(19#2)~#3}
\def\sjnp#1#2#3{{ Sov. J. Nucl. Phys. }{\bf #1~}(19#2)~#3}
\def\jetp#1#2#3{{ JETP Lett. }{\bf #1~}(19#2)~#3}
\def\app#1#2#3{{ Acta Phys. Polon. }{\bf #1~}(19#2)~#3}
\def\rnc#1#2#3{{ Riv. Nuovo Cim. }{\bf #1~}(19#2)~#3}
\def\ap#1#2#3{{ Ann. Phys. }{\bf #1~}(19#2)~#3}
\def\ptp#1#2#3{{ Prog. Theor. Phys. }{\bf #1~}(19#2)~#3}
\def\plb#1#2#3{{ Phys. Lett. }{\bf#1B~}(19#2)~#3}



\newpage

\pagestyle{empty}

\begin{references}

\bibitem{dls} G. Dvali, G. Lazarides and Q. Shafi, 
\pl{424}{98}{259}.

\bibitem{hr} For another approach to MSSM with R-symmetry see
L. J. Hall and L. Randall, \np{352}{91}{289}. 

\bibitem{grs} M. Gell-Mann, P. Ramond and R. Slansky, in 
{\em Supergravity, Proceedings of the Workshop}, Stony Brook, 
New York, 1979, eds. P. Van Nieuwenhuizen and D. Z. Freedman 
(North Holland, Amsterdam, 1979), p. 315; T. Yanagida, in 
{\em Proceedings of the Workshop on Unified Theories and 
Baryon Number in the Universe}, Tsukuba, Japan, 1979, eds. 
A. Sawada and A.Sugamoto (KEK Rep. No. 79-18, Tsukuba, Japan, 1979).

\bibitem{w} C. Wetterich, \np{187}{81}{343}; G. Lazarides, Q. Shafi 
and C. Wetterich, \np{181}{81}{287}; R. N. Mohapatra and G. Senjanovic, 
\pr{23}{81}{165}; R. Holman, G. Lazarides and Q. Shafi, \pr{27}{83}{995}.

\bibitem{ms} E. Ma and U. Sarkar, hep-ph/9802445.

\bibitem{lepto} M. Fukugita and T. Yanagida, \pl{174}{86}{45}; 
W. Buchm\"uller and M. Pl\"umacher, \pl{389}{96}{73}. 
In the context of inflation see G. Lazarides and 
Q. Shafi, \pl{258}{91}{305}; G. Lazarides, ~C. Panagiotakopoulos 
and Q. Shafi, \pl{315}{93}{325}. 

\bibitem{lss} G. Lazarides, R. Schaefer and Q. Shafi, 
\pr{56}{97}{1324}; G. Lazarides, hep-ph/9802415 (to appear in the 
proceedings of the 6th BCSPIN Summer School). 

\bibitem{ibanez} L. E. Ib\'a\~nez and F. Quevedo, \pl{283}{92}{261}.

\bibitem{pq} R. Peccei and H. Quinn, \prl{38}{77}{1440}; 
S. Weinberg, \prl{40}{78}{223}; F. Wilczek, \prl{40}{78}{279}.

\bibitem{lps} G. Lazarides, C. Panagiotakopoulos and Q. Shafi, 
\prl{65}{86}{432}.
  
\bibitem{np} For a recent discussion 
see H. P. Nilles and N. Polonsky, \np{484}{97}{33}.

\bibitem{kn} J. E. Kim and H. P. Nilles, \plb{138}{84}{150}.

\bibitem{dvali} G. Dvali, private communication.

\end{references}
\end{document}